\documentclass[11pt,twocolumn]{article}
\usepackage[utf8]{inputenc}
\usepackage[english]{babel}
\usepackage[numbers]{natbib}
\usepackage[T1]{fontenc}
\usepackage[top=2cm, bottom=2cm, left=1.5cm,right=1.5cm]{geometry}
\usepackage{graphicx}%
\usepackage{xurl}
\usepackage[ruled]{algorithm2e}
\usepackage{algpseudocode}
\graphicspath{{./figures/}}
\usepackage{multirow}%
\usepackage{amsmath,amssymb,amsfonts}%
\usepackage{amsthm}%
\usepackage{mathrsfs}%
\usepackage[title]{appendix}%
\usepackage{xcolor}%
\usepackage{textcomp}%
\usepackage{manyfoot}%
\usepackage{booktabs}%
\usepackage{listings}%
\usepackage{tikz,lipsum,lmodern}
\usepackage{amsmath,amssymb,amsthm}
\newtheorem{lemma}{Lemma}
\newtheorem{theorem}{Theorem}
\theoremstyle{definition}

\usepackage{enumerate}
\usepackage{qcircuit}

\usepackage{caption}
\usepackage{subcaption}
\usepackage[most]{tcolorbox}

\newcommand{\ket}[1]{\ensuremath{\left|#1\right\rangle}} % Dirac Kets
\newcommand{\norm}[1]{\ensuremath{\lVert#1\rVert }}

\title{A quantum compiler design method by using linear combinations of permutations}
\author{Ammar Daskin}
\date{}
\begin{document}
\maketitle

\begin{abstract}
 A matrix can be converted into a doubly stochastic matrix by using two diagonal matrices. And a doubly stochastic matrix can be written as a sum of permutation matrices.
In this paper, we describe a method to write a given generic matrix in terms of quantum gates based on the block encoding.
 In particular, we first show how to convert a matrix into doubly stochastic matrices and by using Birkhoff's algorithm, we express that matrix in terms of a linear combination of permutations which can be mapped to quantum circuits. We then discuss a few optimization techniques that can be applied in a possibly future quantum compiler software based on the method described here.
\end{abstract}

\section{Introduction}

A few recent quantum algorithms such as quantum signal processing \cite{low2017optimal,low2019hamiltonian} and quantum singular value transformation \cite{gilyen2019quantum} operate based on encoding a generic $2^n \times 2^n$ matrix $A$ as a subsystem of larger unitary matrix, $U$:
\begin{equation}
\label{Eq:blockencoding}
    \mathcal{U} = \left(\begin{matrix}
        A&\bullet\\
        \bullet & \bullet\\
    \end{matrix}\right).
\end{equation}
where `` $\bullet$ '' symbols represent matrix parts that are not used in the computations involving application of $A$  to a considered arbitrary $n$-qubit quantum state \ket{\psi}.
In terms of quantum circuits, the system is in general divided into two distinct quantum registers: \ket{ancilla}\ket{\psi}. While the size of register $\ket{\psi}$ is $n$-qubits, the size of the ancilla is determined by the extra qubits used to build $ \mathcal{U}$. 
In the circuit, we start with an initial multi-qubit \ket{ancilla}=\ket{0}.
Then, the application of $U$ to \ket{ancilla}\ket{\psi} generates a quantum state:
\begin{equation}
     \mathcal{U}\ket{ancilla}\ket{\psi} = \ket{0}A\ket{\psi}+\sum_{i, i\neq 0}\ket{i} \bullet \ket{\psi},
\end{equation}
where \ket{i} represents the $i^\text{th}$ vector in the standard basis.
As seen in the above, when \ket{ancilla} = \ket{0}, we have the application $A\ket{\psi}$ in the main register. 

Generating circuit for $ \mathcal{U}$ or using it to estimate eigenvalues of $A$ can be achieved by writing $A$ as a linear combination of unitary matrices \cite{childs2012hamiltonian,daskin2012universal,daskin2014universal,daskin2017ancilla}, using quantum signal processing \cite{low2017optimal,low2019hamiltonian}, or a linear combination of Hamiltonians \cite{an2023linear}.

Most quantum gates can be expressed by using permutation matrices. Vice versa most permutation matrices can be easily implemented by quantum computers. 
Permutations along with generic rotation matrices are also important to factorize a matrix into different forms.
Because of these, many circuit design techniques are based on decomposition of a matrix into simpler forms by using permutation-like matrices. 
For example, circuits for Schur transformations  designed in Ref.\cite{bacon2006efficient} mostly uses permutations. Ref.\cite{fijany1999quantum} also uses them to design circuits for quantum wavelet transforms. 
A quantum circuit transformation is proposed \cite{childs2019circuit} to fit a given transformation into an architecture based on  considering  them on permutation graphs and using routing via matching \cite{alon1993routing} which converts a permutation into the product of disjoint transpositions and token swapping \cite{yamanaka2015swapping} which minimizes the number of transpositions.
Since their graphs are also related to quantum walks, the circuit implementation for quantum walks such as \cite{douglas2009efficient,loke2017efficient}  can be also used to implement circuits for permutations. Permutation matrices are also key ingredients in designing pseudo-random circuits (e.g. \cite{chen2024efficient}. 

In terms of the computational complexity, the permutation matrices can be implemented in $O(poly(n))$, where $n$ is the number of qubits, since they are row and column sparse \cite{berry2007efficient}. A generic circuit equivalent of a permutation matrix can be simply defined by a combination of $X$ and controlled-$X$ gates. 
Therefore, an application of a group of $k$ permutation matrices in most cases can be simplified and further optimized to reduce the complexity of $O(kpoly(n))$.

A matrix is row ( or column) stochastic if its row (or column) sums are 1. Stochastic matrices are used in Markov processes and in many field of science.
A circuit decomposition for \underline{symmetric} stochastic matrices are given in Sec.5 of Ref.\cite{camps2022explicit} based on finding a unitary matrix which involves the given matrix as encoded in  Eq.\eqref{Eq:blockencoding}. 
Ref.\cite{camps2022explicit} in particular have showed that how to implement Chebishev polynomials of a stochastic matrix $W$,  $T_k(W)$ and $T_k(W/\alpha)$ when there is a circuit for $W$ or $W/\alpha$ where $\alpha$ is a real value used for scaling. 
It is then showed how it is connected to Szegedy's quantum walk \cite{szegedy2004quantum} with using a SWAP operator for quantum registers and a walk matrix $W$ that implements transition probabilities.

In this paper, we describe a method to obtain exact circuit implementation of a matrix inside a block encoding circuit by first converting matrix into a doubly stochastic matrix. Then, we write it as a linear combination of permutation matrices. We then show how to map a generic permutation matrix on quantum circuits and implement a doubly stochastic matrix written as a linear combination of permutations. Then, we discuss a few optimization ideas that can be used to reduce the number of quantum gates obtained in the linear combination of permutations and give future directions to implement full-fledged quantum compiler software based on the method described here.

\section{Doubly Stochastic(bistochastic) Matrices}
A matrix $S$ is doubly stochastic if its matrix elements $S_{ij}$ are greater or equal to zero and their row and column sums are equal to one:  i.e. $S_{ij}\geq 0$ and  $\sum_i S_{ij} = \sum_j S_{ij}=1$.

According to Sinkhorn's theorem, a square matrix $A$ with strictly positive elements can be converted into a doubly stochastic matrix $S$ by  multiplication of two diagonal matrices $D_L$ and $D_R$ with strictly positive elements, as follows: 
\begin{equation}
    S = D_LAD_R.
\end{equation}  
Similarly, there exists unitary-diagonal matrices $U_L$ and $U_R$ such that the following matrix is doubly stochastic  whose column and row sums are equal to 1:
\begin{equation}
    U_S=U_LUU_R.
\end{equation}

For a circulant matrix $C$ with the first row elements $ [c_1, c_2\dots, ] $ and their sum $c = \sum c_j$, the scaled circulant matrix, $C/c$,  is also a doubly stochastic matrix \cite{dufosse2016notes}.

We can also convert a matrix by extending its dimensions. For instance, a stochastic matrix $T$ can be converted into doubly stochastic matrix as follows:
\begin{equation}
   S =  \left(\begin{matrix}
        T/\alpha&(1-1/\alpha)I \\
        diag(s)&T^T/\alpha
    \end{matrix}\right),
\end{equation}
where $\alpha$ is the maximum row sum of $T$, $I$ is the identity matrix and $s$ is a diagonal matrix that complements column sums to 1.

The  graph associated with a doubly stochastic matrix has a perfect matching. 
Birkhoff-von Neumann Theorem \cite{birkhoff1946tres} states that a matrix is doubly stochastic if and only if it lies in the convex hull of the set of permutation matrices. In other words, any doubly stochastic matrix $S$ can be written as a convex combination of $k$ permutation matrices:
\begin{equation}
    S = w_1 P_1 + \dots + w_k P_k, \text{ with } 0\leq w_i \leq 1, \text{ and }\sum_i^kw_i = 1.
\end{equation}
Here, $k$ is bounded in Ref.\cite{marcus1959diagonals,dufosse2018further} as $k\leq (N-1)^2+1$ with $N$ being the dimension of the matrix. 
$P_i$s are permutation matrices which are orthogonal: i.e. $PP^T = P^TP = I$, and row and column sparse by containing only single 1 in each column and row.  Their products with each other are also another permutation matrix and  their powers go to an identity matrix with some order. 

This theorem with the associated Birkoff algorithm described below has application in tensor decomposition\cite{chen2019birkhoff},  quantum mechanics \cite{louck1997doubly}, graph problems \cite{brualdi1988some},  optimization \cite{fiedler2022doubly},  qubit routing problem \cite{mariella2023doubly}, and circuit switching \cite{valls2021birkhoff}.

\subsection{Difficulty of Finding This Decomposition}
The decision version of the above problem-i.e. whether there is a decomposition with $k$ permutation matrices-is shown to be NP complete problem \cite{dufosse2016notes,brualdi1982notes}. That means it requires searching through significantly many $k$ group of permutations among the possible $N!$ different permutations. Therefore, taking the above decomposition as an optimization problem is an NP-hard problem. 
However, as stated by Marcus-Ree theorem \cite{marcus1959diagonals,dufosse2018further},  there exist at least one decomposition with the  value $ k\leq (n-1)^2+1$.

 It is also shown that by using a subset of admissible permutation matrices, $\epsilon$-approximate decomposition with $O(log(1/\epsilon))$ number of permutations is possible \cite{kulkarni2017minimum, valls2021birkhoff}.

\subsection{Birkhoff's Algorithm}
 Although the original algorithm is a greedy algorithm and can be found in many different resources (e.g. \cite{dufosse2016notes}), there also a few other variants: a mixed integer linear programming \cite{dufosse2018further,kulkarni2017minimum} which finds a decomposition with the smallest integer $k$ and its mix-max variants \cite{valls2021birkhoff}.
For self-containment of the paper, we give the greedy version of the Birkhoff's algorithm \cite{dufosse2016notes} in Algorithm \ref{Birkhoffs algorithm} and also code in provided data link\footnote{\url{https://github.com/adaskin/stochastic-matrix-into-permutations}}:
 The  algorithm, at each step $i$, finds a perfect matching permutation $P_i$ based on the input matrix $S_i$. In the original algorithm $P_i$ always consists of a 1 associated to the minimum non-zero element of $S_i$. The algorithm computes a coefficient $p_i$ which is the minimum among the elements of $S_i$ corresponding to the 1s in the permutation matrix.
In this algorithm, we generate a permutation matrix for the smallest elements of $S_i$ and then for the smallest of the remaining parts. This way the elements of $S_i$ always remain positive.

\begin{algorithm}
\caption{Birkhoff-von Neumann Algorithm}
\label{Birkhoffs algorithm}
\textbf{Input:} $S, k$ \Comment{a doubly stochastic matrix}\\
$i = 0$\\
$P = \{\}$ \Comment{permutation matrices}\\
$w = [ ]$\Comment{a list of  weights}\\
$S_i = S.copy()$\\
\While{$\norm{S_i} \not\approx 0$ \textbf{and} $i < k$ }{
find a permutation $P_i \subseteq S_i$\\
compute weight $w_i$: ( e.g., $w_i \gets min (P_i \odot S_i$))\\
$S_i = S_i - w_iP_i$\\
$w.append(w_i)$\\
$P[i] = P_i$\\
$i = i + 1$\\
}
\textbf{Output: $P, w, ||S_i||$} 
\end{algorithm}

An example decomposition from the algorithm is illustrated below \cite{brualdi1982notes}:

\begin{equation}
\label{Eq:exampleperms}
\small
\begin{split}
  \underbrace{
    \frac{1}{6}\left(\begin{matrix}
        1&4&0&1\\
        2&1&3&0\\
        2&1&1&2\\
        1&0&2&3
    \end{matrix}\right)}_{S} = &
        \frac{1}{6}
    \underbrace{
        \left(\begin{matrix}
        0&0&0&1\\
        0&0&1&0\\
        0&1&0&0\\
        1&0&0&0
    \end{matrix}\right)}_{P_1}
    +
    \frac{1}{6} \underbrace{
    \left(\begin{matrix}
        1&0&0&0\\
        0&1&0&0\\
        0&0&1&0\\
        0&0&0&1
    \end{matrix}\right)}_{P_2}\\
    &+ 
    \frac{1}{3}  \underbrace{\left(\begin{matrix}
        0&1&0&0\\
        1&0&0&0\\
        0&0&0&1\\
        0&0&1&0
    \end{matrix}\right)}_{P_3}
    +       
    \frac{1}{3} \underbrace{\left(\begin{matrix}
        0&1&0&0\\
        0&0&1&0\\
        1&0&0&0\\
        0&0&0&1
    \end{matrix}\right)}_{P_4}
\end{split}
\end{equation}

\section{Quantum Circuit Implementations}
\subsection{Quantum Circuit of a Permutation Matrix}

A permutation matrix (see e.g. \cite{dixon1996permutation} is the \underline{row} permuted version of an identity matrix whose columns are indexed as $(0, \dots, N-1)$.
The permutation matrix can be represented by indicating row orderings in two or one row notations, e.g.:
\begin{equation}
   t = \left(\begin{matrix}
        0& 1& 2& 3\\
        1& 2& 0& 3
    \end{matrix}\right)=
    \left(\begin{matrix}
        0& 1& 2& 3\\
        t(0)& t(1)& t(2)& t(3)
    \end{matrix}\right)
    = 1203.
\end{equation}
When a permutation matrix is of a considerably small dimension, it is easy to map them into quantum circuits. For instance, we can map the above example decomposition to the quantum gates as follows:
\begin{align}
&P_1 = X\otimes X & 
\Qcircuit @C=1em @R=.7em {
& \gate{X}  & \qw \\
& \gate{X}  & \qw
} \\
&P_2 = I &\Qcircuit @C=1em @R=.7em {
& \qw \\
&  \qw
}\\
&P_3 = I\otimes X& 
\Qcircuit @C=1em @R=.7em {
& \qw &  \qw \\
& \gate{X} & \qw
}\\
&P_4 =C_{\ket{0}}^0X_1\times C_{\ket{0}}^1X_0&
\Qcircuit @C=1em @R=.7em {
& \ctrlo{1} & \targ & \qw \\
& \targ & \ctrlo{-1} & \qw
}
\end{align}
However, for more complex permutations we have to decompose them into smaller terms. Below we go through their compositions and show how to define them in terms of transpositions which can be directly mapped to quantum gates. 

\subsection{Composition and Commutation}
Composition of two permutations is also a permutation matrix. As an example, consider the following permutations:
\begin{align}
   t = \left(\begin{matrix}
        0& 1& 2& 3\\
        1& 2& 0& 3
    \end{matrix}\right),\ 
    s = \left(\begin{matrix}
        0& 1& 2& 3\\
        3& 2& 1&0
    \end{matrix}\right)
\end{align}
The composition $t\circ s$ indicates a permutation $s$, then a permutation $t$ for each row index $i$:
i.e. $t(s(i))$:
\begin{equation}
\begin{split}
    &t(s(0)) = t(3) = 3,\ t(s(1))= t(2) = 0,\\ & t(s(2)) =t(1)= 2,\ t(s(3)) =t(0)= 1, \\
    &\to t\circ s = \left(\begin{matrix}
        0& 1& 2& 3\\
        3& 0& 2&1
    \end{matrix}\right)
    \end{split}
\end{equation}

A permutation can be also written in cyclic form in terms of a disjoint set of cycles. Since the above permutation $t$ goes from 0 to 1, from 1 to 2, and from 2 to the starting point 0, it forms a cycle. Therefore, it can be rewritten as $(012)(3)$. Similarly, $s=(03)(12)$. 

The disjoint cycles commute; that means, they can be written in any order. 
In addition,  the numbers within a cycle can be written starting at any number (written in a circular way): e.g. $(012) = (120) =  (201)$.

The product(composition) of permutations can be also found by simplifying the combined cyclic notation into disjoint cycles.
Note that since the compositions are not commutative in general, in the above example $ts\neq st$, the permutation group is a non-abelian group. However, the disjoint cycles commute, therefore their order in the product can be rearranged: e.g. $(03)(12)=(12)(03)$.

\subsection{Decomposition into Transpositions}
In cyclic notation, the cycles with length two (\emph{transpositions}) represent a swap between states. 
Every permutation can be decomposed as a product of 2-cycles. 
For instance,
$t=(012)(3)$ can be rewritten as $(02)(01)(3)$.
And this decomposition may be written in many different ways: e.g.  $t$ can be also written as $(01)(12)(3)$.
Note that by simply pairing rows of the matrices, we can have at most $N/2$ disjoint transpositions. 

For distinct numbers $a_0,a_1, \dots a_m$, the following decomposition holds:
\begin{equation}
   (a_0,\dots, a_m) = (a_0,a_1)(a_1,a_2,\dots, a_m). 
\end{equation}
Repeating the above recursively, we can describe any cycle in terms of transpositions:
\begin{equation}
\label{eq:cycledecomp1}
\begin{split}
    (a_0,\dots, a_m) = & (a_0,a_1)(a_1,a_2)(a_2,a_3)\\
    &\dots (a_{m-2}, a_{m-1}) (a_{m-1}, a_m).
    \end{split}
\end{equation}
Alternatively, we can decompose a cycle in the following way:
\begin{equation}
\label{eq:cycledecomp1}
    (a_0,\dots, a_m) = (a_0,a_m)(a_0,a_{m-1})\dots (a_{0}, a_{2}) (a_{0}, a_1). 
\end{equation}
Above two equations can be used as algorithms to write any permutations in terms of transpositions.
Here, note that since \underline{disjoint} transpositions commute, the permutations and their products written in this form can be further simplified.

\subsection{Circuit Implementation of a Transposition}
Some of the transpositions can be implemented as a quantum circuit by using only a single multi-controlled $X$ gate.
\begin{theorem}
\label{theorem_transwithXgate}
    If the Hamming distance between $a_i$ and $a_j$ is one, then a transposition $(a_ia_j)$ can be implemented by using a multi-controlled $X$ gate whose target is the qubit where  $a_i$ and $a_j$ are different and control qubits are all the other qubits.
    \begin{proof}
        The proof comes from the definition of the $X$ gate: The target qubit is the only qubit whose state changes. Since the rest of the qubits are the control and only activates $X$ in \ket{a_i} and \ket{a_j},  they will swap and the rest of the states remain unchanged.  
    \end{proof}
\end{theorem}
As an   example consider the cycle (13) in a three-qubit system. We can implement this swap by using an ancilla register as done in Ref.\cite{facchini2015quantum}, however, it would increase the complexity because of the additional controlled registers.
Instead, we first rewrite this in binary form $(13)=(011-111)$. Then, we observe the changing and unchanging qubits: In this example, we have an $X$ gate on the first applied when the second and third qubits are \ket{1}. In terms of quantum circuit, it can be simply depicted as:
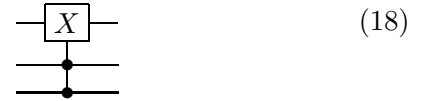
\begin{equation}
    \Qcircuit @C=1em @R=.7em {
&\gate{X} &  \qw \\
& \ctrl{-1}&  \qw\\
& \ctrl{-1}& \qw\\
}
\end{equation}

\begin{lemma}
\label{lemma_decomoftransposition}
    If the Hamming distance between $a_i$ and $a_j$ is $h>1$,  we can decompose $ (a_ia_j)$ as follows:
    \begin{equation}
        (a_ia_j) = (a_i b_1)(b_1b_2)\dots (b_{h-1}b_h)(b_ha_j),
    \end{equation}
    where the Hamming distance between $a_i$ and $b_1$,  any $b_i$ and $b_{i+1}$, and $b_h$ and $a_j$ is one.
\end{lemma}
\begin{theorem}
    If the Hamming distance between $a_i$ and $a_j$ is $h$, then a transposition $(a_ia_j)$ can be implemented by using $h$ number of multi-controlled $X$ gates.
    \begin{proof}
        It follows the decomposition given in the above Lemma \ref{lemma_decomoftransposition} along with Theorem \ref{theorem_transwithXgate}. 
    \end{proof}
\end{theorem}

The above implementation for a single transposition can be
combined to implement any permutation matrix written in
terms of transpositions even though it may consist of many redundancies. 
As mentioned before,  an optimized implementation can be also found in Ref. \cite{childs2019circuit} which
minimizes the number of transpositions by using token swapping\cite{yamanaka2015swapping}. 

\subsection{Circuit Implementation of a Doubly Stochastic Matrix}
After writing $S$ as a linear combination of permutation matrices, we can simply construct a circuit for it by using an ancilla register and controlling permutation matrices (which are recently called iteration circuits e.g. \cite{ruiz2024quantum}).
For instance, we can map the linear combination of permutations given in Eq.\eqref{Eq:exampleperms} as follows:
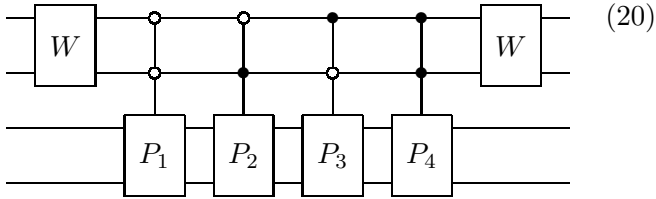
\begin{equation}
    \Qcircuit @C=1em @R=1em {
 &\multigate{1}{W}& \ctrlo{1}&\ctrlo{1}&\ctrl{1}&\ctrl{1}&\multigate{1}{W}&\qw\\
 &\ghost{W}& \ctrlo{1}&\ctrl{1}&\ctrlo{1}&\ctrl{1}&\ghost{W}&\qw\\
 &\qw& \multigate{1}{P_1}
    &\multigate{1}{P_2}&\multigate{1}{P_3}&\multigate{1}{P_4}&\qw&\qw\\
&\qw&\ghost{P_1} &\ghost{P_2} &\ghost{P_1} &\ghost{P_2} &\qw&\qw
} 
\end{equation}
Here, $W$ implements the coefficients in the linear combination: It essentially converts \ket{0} ancillary input state into a superposition state with the square root of the normalized coefficients, i.e. $\sqrt{w_i}$s.

\section{Discussion and Future Directions}

\subsection{Simplifications of Permutations and their Combinations}
After writing a matrix as a product of permutations represented in the forms of disjoint cycles, many string algorithms along with specific data structures such as suffix arrays or graphs can be used to find similarities between permutations and simplify them.
Two permutations are similar if they have the same cycle decomposition. Therefore, a product and/or sum of permutations can be simplified by observing their similar parts.
Here, to simplify circuit design, the following observations can be used to design optimization parts of a compiler:
\begin{enumerate}
    \item \textbf{Reduction:} The first observation is that if the same gate is applied in different cycles with different states of the same control qubits, then these gates can be simplified. 
For instance, in cycles $(011-111)(001-101)$ written in binary form, we apply $X$ to the first qubit no matter what the state of the second qubit is. This permutation is simply equivalent to the following:
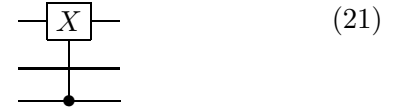
\begin{equation}
    \Qcircuit @C=1em @R=1em {
&\gate{X} &  \qw \\
& \qw&  \qw\\
& \ctrl{-2}& \qw\\
}
\end{equation}

\item \textbf{Grouping:} The second observation is that if they have common control qubits which are not target qubits as shown below, we can still group them into a single permutation controlled by the common control qubits.
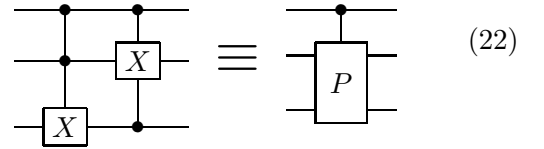
\begin{equation}
    \begin{aligned}
    \Qcircuit @C=1em @R=1em {
& \ctrl{1}&  \ctrl{1}&\qw\\
& \ctrl{1}& \gate{X}&\qw\\
&\gate{X} &  \ctrl{-1} &\qw\\
}& \begin{matrix}
    &\\ &\\ &\\
   & \scalebox{2}{$\equiv$}
\end{matrix}&
    \Qcircuit @C=1em @R=1em {
& \ctrl{1}&\qw\\
& \multigate{1}{P}&\qw\\
&\ghost{P} &\qw
}
\end{aligned}
\end{equation}
This can help to eliminate the operations whose combinations are the same as the other group of operations.
\item \textbf{Ungrouping:} The reverse operation of the grouping.

\item  \textbf{Reordering:} The last observation is that in addition to reordering of general disjoint cycles (they commute);  the gates can be reordered with different control bit values without changing their relative order to the gates with the same control values. This can also help grouping or simplifying permutations by allowing to move the gates in time on the circuit. An example of reordering is given below:
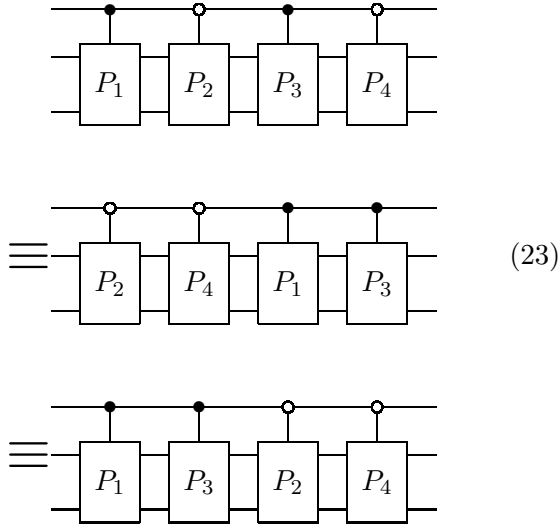
\begin{equation}
    \begin{aligned}&
    \Qcircuit @C=1em @R=1em {
& \ctrl{1}&\ctrlo{1}&\ctrl{1}&\ctrlo{1}&\qw\\
& \multigate{1}{P_1}&\multigate{1}{P_2}&\multigate{1}{P_3}&\multigate{1}{P_4}&\qw\\
&\ghost{P_1} &\ghost{P_2} &\ghost{P_1} &\ghost{P_2} &\qw
} & \\ 
\begin{matrix}
    &\\ &\\ &\\
   & \scalebox{2}{$\equiv$}
\end{matrix} &
    \Qcircuit @C=1em @R=1em {
& \ctrlo{1}&\ctrlo{1}&\ctrl{1}&\ctrl{1}&\qw\\
& \multigate{1}{P_2}&\multigate{1}{P_4}&\multigate{1}{P_1}&\multigate{1}{P_3}&\qw\\
&\ghost{P_2} &\ghost{P_1} &\ghost{P_2} &\ghost{P_1} &\qw
}
& \\ \begin{matrix}
    &\\ &\\ &\\
   & \scalebox{2}{$\equiv$}
\end{matrix} &
    \Qcircuit @C=1em @R=1em {
& \ctrl{1}&\ctrl{1}&\ctrlo{1}&\ctrlo{1}&\qw\\
& \multigate{1}{P_1}&\multigate{1}{P_3}&\multigate{1}{P_2}&\multigate{1}{P_4}&\qw\\
&\ghost{P_2} &\ghost{P_1} &\ghost{P_2} &\ghost{P_1} &\qw
}
\end{aligned}
\end{equation}

\end{enumerate}

\subsection{Approximation by Cutting off the Terms in the Summation}
A matrix $S$ can be approximated by removing permutations with smaller weights from the corresponding linear combination.
In this case since $\sum_i w_i = 1$, the approximation error is directly related to the sum of weights that are ignored.

\subsection{Quantum Compiler Design}
A classical compiler design \cite{aho1977principles,grune2012modern} in general includes lexical and semantic analyses, intermediate representation, generic optimization, architecture specific optimization, and code generation including register allocations.
And a compiler is in general designed to convert a code in programming language into machine language which is defined by some basic operations(the instruction set) for a classical processor including arithmetic, logic, and control operations.

The same principles can be applied to quantum computers. However, some operations are more expensive and  cannot be easily defined for quantum operations since the current quantum computers are yet to have a functional quantum memory (in classical terms, not a von Neumann machine.). 
The permutations written in cycle forms define quantum operations and can be integrated into a  quantum compiler software and a quantum programming language abstraction can be defined through permutation strings for which string algorithms, such as lexical, semantic, and language processing algorithms, can be applied. 
Therefore, the circuit method described in this paper can help to design quantum programming languages and compilers.

\subsection{Hamiltonian Simulation}
Quantum Hamiltonians can be written in terms of Pauli spin operators (e.g. Ising Hamiltonian which is used for NP problems \cite{lucas2014ising}) or  annihilation and creation operators. Since any of these terms can be converted into a sum of permutations, one can write a Hamiltonian as a linear combination of permutations with complex weights, $w_j \in \mathbb C$. 

One can also sample from a linear combination of permutations by choosing different non-zero sets of weights.

\section{Conclusion}
In this paper, we have shown how to implement a doubly stochastic matrix by writing it as a linear combination of permutations. 
Since any matrix can be converted into doubly stochastic matrices by using diagonal matrices, the method described here can be applied to non-stochastic matrices.

Permutations can be represented by single line strings in memory. 
We have shown that a few observations can be used along with string optimization techniques to simplify permutations and their combinations when generating equivalent quantum circuits. The method described here is generic and we believe it can be integrated into a quantum compiler software or similar algorithmic tools.

\bibliographystyle{unsrtnat}
\bibliography{paper}

\begin{thebibliography}{37}
\providecommand{\natexlab}[1]{#1}
\providecommand{\url}[1]{\texttt{#1}}
\expandafter\ifx\csname urlstyle\endcsname\relax
  \providecommand{\doi}[1]{doi: #1}\else
  \providecommand{\doi}{doi: \begingroup \urlstyle{rm}\Url}\fi

\bibitem[Low and Chuang(2017)]{low2017optimal}
Guang~Hao Low and Isaac~L Chuang.
\newblock Optimal hamiltonian simulation by quantum signal processing.
\newblock \emph{Physical review letters}, 118\penalty0 (1):\penalty0 010501, 2017.

\bibitem[Low and Chuang(2019)]{low2019hamiltonian}
Guang~Hao Low and Isaac~L Chuang.
\newblock Hamiltonian simulation by qubitization.
\newblock \emph{Quantum}, 3:\penalty0 163, 2019.

\bibitem[Gily{\'e}n et~al.(2019)Gily{\'e}n, Su, Low, and Wiebe]{gilyen2019quantum}
Andr{\'a}s Gily{\'e}n, Yuan Su, Guang~Hao Low, and Nathan Wiebe.
\newblock Quantum singular value transformation and beyond: exponential improvements for quantum matrix arithmetics.
\newblock In \emph{Proceedings of the 51st Annual ACM SIGACT Symposium on Theory of Computing}, pages 193--204, 2019.

\bibitem[Childs and Wiebe(2012)]{childs2012hamiltonian}
Andrew~M Childs and Nathan Wiebe.
\newblock Hamiltonian simulation using linear combinations of unitary operations.
\newblock \emph{arXiv preprint arXiv:1202.5822}, 2012.

\bibitem[Daskin et~al.(2012)Daskin, Grama, Kollias, and Kais]{daskin2012universal}
Anmer Daskin, Ananth Grama, Giorgos Kollias, and Sabre Kais.
\newblock Universal programmable quantum circuit schemes to emulate an operator.
\newblock \emph{The Journal of chemical physics}, 137\penalty0 (23), 2012.

\bibitem[Daskin et~al.(2014)Daskin, Grama, and Kais]{daskin2014universal}
Anmer Daskin, Ananth Grama, and Sabre Kais.
\newblock A universal quantum circuit scheme for finding complex eigenvalues.
\newblock \emph{Quantum information processing}, 13:\penalty0 333--353, 2014.

\bibitem[Daskin and Kais(2017)]{daskin2017ancilla}
Ammar Daskin and Sabre Kais.
\newblock An ancilla-based quantum simulation framework for non-unitary matrices.
\newblock \emph{Quantum Information Processing}, 16:\penalty0 1--17, 2017.

\bibitem[An et~al.(2023)An, Liu, and Lin]{an2023linear}
Dong An, Jin-Peng Liu, and Lin Lin.
\newblock Linear combination of hamiltonian simulation for nonunitary dynamics with optimal state preparation cost.
\newblock \emph{Physical Review Letters}, 131\penalty0 (15):\penalty0 150603, 2023.

\bibitem[Bacon et~al.(2006)Bacon, Chuang, and Harrow]{bacon2006efficient}
Dave Bacon, Isaac~L Chuang, and Aram~W Harrow.
\newblock Efficient quantum circuits for schur and clebsch-gordan transforms.
\newblock \emph{Physical review letters}, 97\penalty0 (17):\penalty0 170502, 2006.

\bibitem[Fijany and Williams(1999)]{fijany1999quantum}
Amir Fijany and Colin~P Williams.
\newblock Quantum wavelet transforms: Fast algorithms and complete circuits.
\newblock In \emph{Quantum Computing and Quantum Communications: First NASA International Conference, QCQC’98 Palm Springs, California, USA February 17--20, 1998 Selected Papers}, pages 10--33. Springer, 1999.

\bibitem[Childs et~al.(2019)Childs, Schoute, and Unsal]{childs2019circuit}
Andrew~M Childs, Eddie Schoute, and Cem~M Unsal.
\newblock Circuit transformations for quantum architectures.
\newblock \emph{arXiv preprint arXiv:1902.09102}, 2019.

\bibitem[Alon et~al.(1993)Alon, Chung, and Graham]{alon1993routing}
Noga Alon, Fan~RK Chung, and Ronald~L Graham.
\newblock Routing permutations on graphs via matchings.
\newblock In \emph{Proceedings of the twenty-fifth annual ACM symposium on Theory of Computing}, pages 583--591, 1993.

\bibitem[Yamanaka et~al.(2015)Yamanaka, Demaine, Ito, Kawahara, Kiyomi, Okamoto, Saitoh, Suzuki, Uchizawa, and Uno]{yamanaka2015swapping}
Katsuhisa Yamanaka, Erik~D Demaine, Takehiro Ito, Jun Kawahara, Masashi Kiyomi, Yoshio Okamoto, Toshiki Saitoh, Akira Suzuki, Kei Uchizawa, and Takeaki Uno.
\newblock Swapping labeled tokens on graphs.
\newblock \emph{Theoretical Computer Science}, 586:\penalty0 81--94, 2015.

\bibitem[Douglas and Wang(2009)]{douglas2009efficient}
BL~Douglas and JB~Wang.
\newblock Efficient quantum circuit implementation of quantum walks.
\newblock \emph{Physical Review A}, 79\penalty0 (5):\penalty0 052335, 2009.

\bibitem[Loke and Wang(2017)]{loke2017efficient}
Thomas Loke and Jingbo~B Wang.
\newblock Efficient quantum circuits for szegedy quantum walks.
\newblock \emph{Annals of Physics}, 382:\penalty0 64--84, 2017.

\bibitem[Chen et~al.(2024)Chen, Bouland, Brand{\~a}o, Docter, Hayden, and Xu]{chen2024efficient}
Chi-Fang Chen, Adam Bouland, Fernando~GSL Brand{\~a}o, Jordan Docter, Patrick Hayden, and Michelle Xu.
\newblock Efficient unitary designs and pseudorandom unitaries from permutations.
\newblock \emph{arXiv preprint arXiv:2404.16751}, 2024.

\bibitem[Berry et~al.(2007)Berry, Ahokas, Cleve, and Sanders]{berry2007efficient}
Dominic~W Berry, Graeme Ahokas, Richard Cleve, and Barry~C Sanders.
\newblock Efficient quantum algorithms for simulating sparse hamiltonians.
\newblock \emph{Communications in Mathematical Physics}, 270:\penalty0 359--371, 2007.

\bibitem[Camps et~al.(2022)Camps, Lin, Van~Beeumen, and Yang]{camps2022explicit}
Daan Camps, Lin Lin, Roel Van~Beeumen, and Chao Yang.
\newblock Explicit quantum circuits for block encodings of certain sparse matrices.
\newblock \emph{arXiv preprint arXiv:2203.10236}, 2022.

\bibitem[Szegedy(2004)]{szegedy2004quantum}
Mario Szegedy.
\newblock Quantum speed-up of markov chain based algorithms.
\newblock In \emph{45th Annual IEEE symposium on foundations of computer science}, pages 32--41. IEEE, 2004.

\bibitem[Dufoss{\'e} and U{\c{c}}ar(2016)]{dufosse2016notes}
Fanny Dufoss{\'e} and Bora U{\c{c}}ar.
\newblock Notes on birkhoff--von neumann decomposition of doubly stochastic matrices.
\newblock \emph{Linear Algebra and its Applications}, 497:\penalty0 108--115, 2016.

\bibitem[Birkhoff(1946)]{birkhoff1946tres}
Garrett Birkhoff.
\newblock Tres observaciones sobre el algebra lineal.
\newblock \emph{Univ. Nac. Tucuman, Ser. A}, 5:\penalty0 147--154, 1946.

\bibitem[Marcus and Ree(1959)]{marcus1959diagonals}
Marvin Marcus and Rimhak Ree.
\newblock Diagonals of doubly stochastic matrices.
\newblock \emph{The Quarterly Journal of Mathematics}, 10\penalty0 (1):\penalty0 296--302, 1959.

\bibitem[Dufoss{\'e} et~al.(2018)Dufoss{\'e}, Kaya, Panagiotas, and U{\c{c}}ar]{dufosse2018further}
Fanny Dufoss{\'e}, Kamer Kaya, Ioannis Panagiotas, and Bora U{\c{c}}ar.
\newblock Further notes on birkhoff--von neumann decomposition of doubly stochastic matrices.
\newblock \emph{Linear Algebra and its Applications}, 554:\penalty0 68--78, 2018.

\bibitem[Chen et~al.(2019)Chen, Qi, Caccetta, and Zhou]{chen2019birkhoff}
Haibin Chen, Liqun Qi, Louis Caccetta, and Guanglu Zhou.
\newblock Birkhoff-von neumann theorem and decomposition for doubly stochastic tensors.
\newblock \emph{Linear Algebra and its Applications}, 583:\penalty0 119--133, 2019.

\bibitem[Louck(1997)]{louck1997doubly}
James~D Louck.
\newblock Doubly stochastic matrices in quantum mechanics.
\newblock \emph{Foundations of Physics}, 27:\penalty0 1085--1104, 1997.

\bibitem[Brualdi(1988)]{brualdi1988some}
Richard~A Brualdi.
\newblock Some applications of doubly stochastic matrices.
\newblock \emph{Linear Algebra and its Applications}, 107:\penalty0 77--100, 1988.

\bibitem[Fiedler(2022)]{fiedler2022doubly}
Miroslav Fiedler.
\newblock Doubly stochastic matrices and optimization.
\newblock \emph{by J{\"u}rgen Guddat, Bernd Bank, Horst Hollatz, Peter Kall, Diethard Klatte, Bernd Kummer, Klaus Lommatzsch, Klaus Tammer, Milan Vlach, and Karel Zimmermann, Math. Res}, 45:\penalty0 44--51, 2022.

\bibitem[Mariella and Zhuk(2023)]{mariella2023doubly}
Nicola Mariella and Sergiy Zhuk.
\newblock A doubly stochastic matrices-based approach to optimal qubit routing.
\newblock \emph{Quantum Information Processing}, 22\penalty0 (7):\penalty0 264, 2023.

\bibitem[Valls et~al.(2021)Valls, Iosifidis, and Tassiulas]{valls2021birkhoff}
V{\'\i}ctor Valls, George Iosifidis, and Leandros Tassiulas.
\newblock Birkhoff’s decomposition revisited: Sparse scheduling for high-speed circuit switches.
\newblock \emph{IEEE/ACM Transactions on Networking}, 29\penalty0 (6):\penalty0 2399--2412, 2021.

\bibitem[Brualdi(1982)]{brualdi1982notes}
Richard~A Brualdi.
\newblock Notes on the birkhoff algorithm for doubly stochastic matrices.
\newblock \emph{Canadian Mathematical Bulletin}, 25\penalty0 (2):\penalty0 191--199, 1982.

\bibitem[Kulkarni et~al.(2017)Kulkarni, Lee, and Singh]{kulkarni2017minimum}
Janardhan Kulkarni, Euiwoong Lee, and Mohit Singh.
\newblock Minimum birkhoff-von neumann decomposition.
\newblock In \emph{International Conference on Integer Programming and Combinatorial Optimization}, pages 343--354. Springer, 2017.

\bibitem[Dixon and Mortimer(1996)]{dixon1996permutation}
John~D Dixon and Brian Mortimer.
\newblock \emph{Permutation groups}, volume 163.
\newblock Springer Science \& Business Media, 1996.

\bibitem[Facchini and Perdrix(2015)]{facchini2015quantum}
Stefano Facchini and Simon Perdrix.
\newblock Quantum circuits for the unitary permutation problem.
\newblock In \emph{International Conference on Theory and Applications of Models of Computation}, pages 324--331. Springer, 2015.

\bibitem[Ruiz et~al.(2024)Ruiz, Laakkonen, Bausch, Balog, Barekatain, Heras, Novikov, Fitzpatrick, Romera-Paredes, van~de Wetering, et~al.]{ruiz2024quantum}
Francisco~JR Ruiz, Tuomas Laakkonen, Johannes Bausch, Matej Balog, Mohammadamin Barekatain, Francisco~JH Heras, Alexander Novikov, Nathan Fitzpatrick, Bernardino Romera-Paredes, John van~de Wetering, et~al.
\newblock Quantum circuit optimization with alphatensor.
\newblock \emph{arXiv preprint arXiv:2402.14396}, 2024.

\bibitem[Aho and Ullman(1977)]{aho1977principles}
Alfred~V Aho and Jeffrey~D Ullman.
\newblock \emph{Principles of Compiler Design (Addison-Wesley series in computer science and information processing)}.
\newblock Addison-Wesley Longman Publishing Co., Inc., 1977.

\bibitem[Grune et~al.(2012)Grune, Van~Reeuwijk, Bal, Jacobs, and Langendoen]{grune2012modern}
Dick Grune, Kees Van~Reeuwijk, Henri~E Bal, Ceriel~JH Jacobs, and Koen Langendoen.
\newblock \emph{Modern compiler design}.
\newblock Springer Science \& Business Media, 2012.

\bibitem[Lucas(2014)]{lucas2014ising}
Andrew Lucas.
\newblock Ising formulations of many np problems.
\newblock \emph{Frontiers in physics}, 2:\penalty0 74887, 2014.

\end{thebibliography}

\end{document}